\documentclass[aps,prb,twocolumn,superscriptaddress]{revtex4}
\usepackage{graphicx} %graphics handler
\usepackage{dcolumn}% Align table columns on decimal point
\usepackage{bm}% bold math
\usepackage{amsmath}
\usepackage{epsfig}
\begin{document}

\title{Ultrafast Probes of Nonequilibrium Hole Spin Relaxation \nobreak{in Ferromagnetic Semiconductor GaMnAs}}

%The spin-relaxation time determines how long a nonequilibrium spin polarization can persist

\author{A. Patz}
\affiliation{Ames Laboratory - USDOE and Department of Physics and Astronomy, Iowa State
University, Ames, Iowa 50011, USA}
\author{T. Li}
\affiliation{Ames Laboratory - USDOE and Department of Physics and Astronomy, Iowa State
University, Ames, Iowa 50011, USA}
\author{X. Liu}
\affiliation{Department of Physics, University of Notre Dame, Notre Dame, Indiana 46556, USA}
\author{J. K. Furdyna}
\affiliation{Department of Physics, University of Notre Dame, Notre Dame, Indiana 46556, USA}
\author{I. E. Perakis}
\affiliation{Department of Physics, University of Crete and
Institute of Electronic Structure \& Laser, Foundation for
Research and Technology-Hellas, Heraklion, Crete, 71110, Greece}
\author{J. Wang}
\affiliation{Ames Laboratory - USDOE and Department of Physics and Astronomy, Iowa State
University, Ames, Iowa 50011, USA}

\date{\today}

\begin{abstract}
We directly measure the hole spin lifetime in ferromagnetic GaMnAs via time- and polarization-resolved spectroscopy. Below the Curie temperature T$_{\text{C}}$, an ultrafast photoexcitation with {\em linearly-polarized} light is shown to create a non-equilibrium hole spin population via the dynamical polarization of holes through $p$-$d$ exchange scattering with {\em ferromagnetically-ordered} Mn spins, and we characterize their relaxation dynamics. 
The observed relaxation consists of a distinct three-step recovery : (i) femtosecond (fs) hole spin relaxation $\tau_{S}\sim$160-200 fs, (ii) picosecond (ps) hole energy relaxation $\tau_{E}\sim$1-2 ps, and (iii) a coherent, damped Mn spin precession with a period of $\sim$250 ps. 
The transient amplitude of the $\tau_{S}$ spin component diminishes with increasing temperature, directly following the ferromagnetic order, while the $\tau_{E}$ amplitude shows negligible temperature change, consistent with our interpretation. 
Our results thus establish the hole spin lifetimes in ferromagnetic semiconductors and demonstrate a novel spectroscopy method for studying non-equilibrium hole spins in the presence of correlation and magnetic order.  
%We attribute the short hole spin-relaxation time in GaMnAs to strong spin-orbit interaction in %the valence band mediated by ultrafast momentum scattering in the excited states.  
\end{abstract}
% insert suggested PACS numbers in braces on next line
%\pacs{78.67.Ch,71.35.Ji,78.55.-m}
% insert suggested keywords - APS authors don't need to do this
%\keywords{}

%\maketitle must follow title, authors, abstract, \pacs, and \keywords
\maketitle

\section{Introduction}
Spin correlation, fluctuation and relaxation play important roles in various collective behaviors and macroscopic orders emerging in advanced materials with scientific interest and technological potential, e.g., carrier-mediated ferromagnetism in semiconductors, colossal magnetoresistance in manganites and electronic nematicity in iron pnictide superconductors   \cite{Ohno_APL1996,Li_Nature2013,Patz_NCOMM2014}. 
Since they develop on ultrafast time scales of the order of the inverse magnetic exchange energy, correlated spins can be driven and probed by ultrashort laser pulses interacting with magnetic materials. 
Revealing the associated nonequilibrium spin dynamics may provide some key information, beyond time-averaged properties obtained from static measurements, to both understand and control these phenomena. 

%Despite the potential for much technological importance, an understanding of the hole spin dynamics in semiconductors is still an outstanding issue. 
Recently, nonequilibrium hole spin dynamics in semiconductors has emerged as an outstanding issue. 
For example, in $p$-type doped GaAs, the spin-polarized holes photoexcited by circularly polarized mid-infrared pulses were shown to exhibit an ultrafast exponential decay associated with the hole-spins, and the spin lifetime was found to be $\sim$110 fs at room temperature\cite{Hilton_PRL02}. In contrast, a substantially longer hole spin relaxation time, on the order of hundreds of picoseconds (ps), was inferred from spin tunneling and transport experiments in ferromagnet $p$-Si and $p$-Ge heterostructures \cite{Saroj_Nature, Iba_APE, Shikoh_PRL2013}. Furthermore, in bulk Ge, hole spin lifetimes are reported to differ by two orders of magnitude (700fs vs. 100ps) although some reconciliation was found in that the hole-spin relaxation rate decreased with lower temperature and/or excitation density \cite{Hautmann_PRB2011,Loren_PRB2011}. A unified understanding of the wide range of hole spin lifetimes observed in semiconductors thus far is still missing, and more measurements are clearly desirable in complementary systems, including magnetically-doped semiconductors. A comprehensive and reliable knowledge of the hole spin lifetime is also important for the development of spin computation and communication technology based on conducting holes in conjunction with electrons, which has evolved as an important strategy for implementing spintronics applications \cite{Chappert_Nat2007, Awschalom_Nature07, Zutic_JPCM2007}.

So far, hole spin relaxation has only been studied in weakly-interacting spin ensembles without long-range order, e.g., $p$-GaAs, $p$-Si, $p$-Ge. However, (III,Mn)V magnetic semiconductors such as GaMnAs displaying carrier-mediated ferromagnetism represent a model system for investigating hole spins influenced by ferromagnetic order and spin correlation. 
For example, the exchange interaction between impurity band holes from Mn-doping and localized Mn spins in GaMnAs strongly depends on the hole density, polarization and distribution among the bands \cite{Dobrowolska_Nature2012}. 
Ultrafast spin dynamics and magnetization control in (III,Mn)V magnetic semiconductors is becoming a well-established field and the work presented here was motivated by numerous previous studies on ultrafast cooperative spin phenomena. For example, fs Mn spin canting induced by spin-orbit torques \cite{wang-2009,nemec-2013}  via photo-excited non-thermal ``transverse" hole spins photoexcited due to the interplay between spin--orbit and magnetic exchange interaction \cite{kapetanakis-2009,kapetanakis-2011}, fs demagnetization (decrease in Mn spin amplitude) via dynamical polarization of ``longitudinal" holes spins \cite{Wang_PRL05, Cywinski_PRB07, Wang_JPC06}; ps photoinduced ferromagnetism \cite{Wang_PRL07}, and magnetization precession \cite{Qi_PRB2009, Qi_APL2007, Rozkotova_APL2008, Scherbakov_PRL2010, Zhu_APL2009,  Hashimoto_PRL2008, Rozkotova_APL2008b}.
However, despite extensive studies in these (III,Mn)V ferromagnetic semiconductors, the hole spin relaxation time is yet to be determined and the distinction between the non-equilibrium hole spin and hole energy relaxation are still unknown until this work. In this case, the lack of conclusive observations is due in part to the relatively scarce experimental techniques that can generate and probe spin-polarized holes at femtosecond time scales, especially in heavily hole-doped materials. 
%Moreover, the hole spin dynamics in these (III,Mn)V ferromagnetic materials are still unknown and the hole spin relaxation time is yet to be determined. Ultrafast magnetization control in magnetic semiconductors is a quickly growing field and the work presented here was largely motivated by previous studies on ultrafast demagnetization, magnetization enhancement, and coherent control of spins in magnetic semiconductors \cite{Cywinski_PRB07, Wang_JPC06, Koopmans_PRL00, Qi_PRB2009, Qi_APL2007, Rozkotova_APL2008, Qi_PSSC2008, Scherbakov_PRL2010, Zhu_APL2009, Kapetanakis_PRL2009, Wang_PRB08, Wang_SM2007, Oiwa_JS2005, Hashimoto_PRL2008, Rozkotova_APL2008b, Ohno_APL1996, Wang_PRL05, Wang_PRL07, Astakhov_APL2005}. 

Here, we report the observation of femtosecond hole spin relaxation in ferromagnetic GaMnAs using degenerate ultrafast magneto-optical Kerr (MOKE) spectroscopy. We reveal a femtosecond demagnetization followed by a fast (characteristic time $\tau_{S}$) and a slow (characteristic time $\tau_{E}$) recovery of transient MOKE signals. The fast recovery appears only in the magnetically ordered state below T$_{\text{C}}$ and is characterized by a temperature-dependent transient amplitude that closely follows the trend of the ferromagnetic order. In contrast, the slow recovery persists in the paramagnetic phase and its transient amplitude shows negligible temperature dependence. This temperature dependence suggests that the optical nonlinearity of the $\tau_{S}$ component is determined by the dynamical polarization of hole spins scattering with the magnetically-ordered Mn ion spins through $p$-$d$ exchange scattering. 

We infer from this observation that the dynamics due to the relaxation of the non-equilibrium spin-polarized holes occurs within a $\sim$160-200 fs time interval. This $\tau_S$ component is distinguished in time from the subsequent magneto-optical signal decay characterized by $\tau_{E}\sim$1-2ps, which we attribute to the hole energy relaxation. On longer timescales, a coherent damped Mn spin precession is observed with a period of $\sim$250 ps at zero B field. Our results reveal the values of the hole spin lifetimes and different stages of spin dynamics in magnetic semiconductors. We note that the use of near-infrared, linearily polarized pulses presents a new approach for creating and studying the non-equilibrium hole spins in the presence of ferromagnetic order, which is much simpler to implement in comparison to the mid-infrared, circularly polarized pulses used previously \cite{Hilton_PRL02, Hautmann_PRB2011,Loren_PRB2011, Wang_PRB08, Wang_PRL05}.  

\section{Experimental Schematics and Sample Details}
The photoinduced magnetization dynamics of our GaMnAs sample was studied via ultrafast MOKE spectroscopy. A 1 KHz Ti:Sapphire laser amplifier was used to generate pulses with central wavelengths of 800nm and pulse durations of 35fs. The output pulses were separated into two beam paths for the pump and probe. 
%The pump fluence was 0.69mJ/cm$^{2}$, the probe was set at 0.16\;mJ/cm^{2}$, 
Both beams were linearly polarized. A chopper modulated the pump beam at a frequency of 500Hz. Data was collected through a polarization bridge, balanced photodetector, and lock-in amplifier. In the polar geometry used here, the Kerr rotation angle, $\Delta\theta$, upon reflection from the sample is proportional to the out-of-plane magnetization M$_{z}$ of the sample.
%A noise floor below $4\times10^{-7}$ deg was achieved.

Our sample was a 70nm Ga$_{0.925}$Mn$_{0.075}$As (GaMnAs) thin film, which was deposited on a GaAs buffer layer on top of a semi-insulating $\left[100\right]$ surface of a GaAs substrate by low-temperature molecular beam epitaxy (MBE). The Curie temperature 
and hole density were 77 K and $3 \times 10^{20}$ cm$^{-3}$, respectively. 
The studies were performed in a liquid helium cooled cryostat. 
At low temperature the spontaneous magnetization from the carrier-mediated ferromagnetic order naturally aligns along one of the two equivalent, orthogonal easy axes lying in the sample plane close to the [100] and [010] crystallographic axes. 
%The axes directions are associated with an effective internal magnetic field referred to as the anisotropy field. 
Increasing the temperature above a reorientation temperature, T$_R \sim30$K, the anisotropy field switches direction and now points close to [1$\bar{1}$0] as evidenced by the crossing in Fig. 1(a), which shows magnetizations along four different axes. 
%The magnetization quickly diminishes approaching the critical temperature T$_C \sim77$K. For temperatures, $T > T_R$, the magnetization measured along different directions mostly scales together. 
In our experiment, an external magnetic field, B$_{\text{ext}} \sim250mT$, is applied along the $z$-axis, [001] (inset, Fig. 1(b)), and therefore, the resultant magnetization direction is a resultant of the external and internal ansiotropy fields. A static MOKE measurement in polar geometry as a function of external magnetic field amplitude (along the [001] axis) is shown in Fig. 1(b). 
An increasing MOKE signal is observed with applied external B-field up to $\sim$500mT where the magnetization saturates along the [001] direction at 4K. 

\begin{figure} 
\includegraphics[scale=.13]{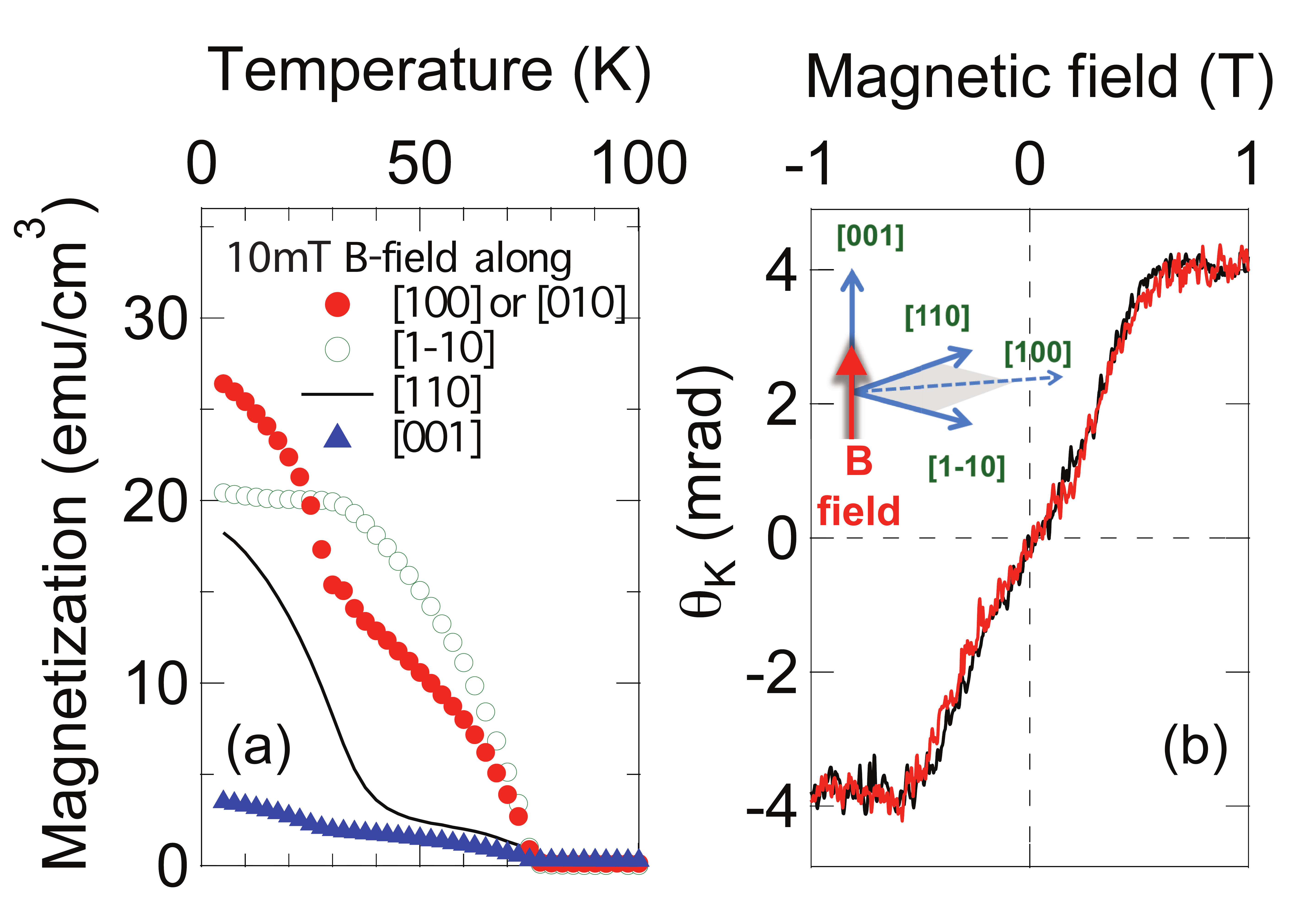} 
\caption{\label{fig1} (a) Static magnetizations at 5K along four different axes [100] or [010] (red solid circle), [1$\bar{1}$0] (green hollow circle), [110] (black line) and [001] (blue triangle). B$_{\text{ext}} \sim 10\text{mT}$ (b) Static MOKE angle, $\theta_k$, measured in polar geometry with external magnetization direction along the Z-axis [001] at 5K. Inset: sample schematics. An external magnetic B field is applied along the $Z$-axis}
\end{figure}

Fig. 2 illustrates the origins of the static and ultrafast MOKE signals in GaMnAs semiconductors for T$<$T$_C$.  The linearly polarized probe beam with energy centered at 1.55eV strongly couples to the electronic transition across the direct band-gap at the $\Gamma$-point.
%The spectral range of the pulses covers 1.59-1.53eV which are all greater then the band-gap for all temperatures.
The valance bands
%(heavy-hole, light-hole, and split-off bands)
are spin-split due to the strong exchange coupling with the ferromagnetically aligned Mn spins in $d$-like states (shown in gold) that hybridize with the upper valence band.
The size of the splitting is proportional to the magnetization from Mn ions and forms the mean-field gap, $\Delta_{MF}$, as shown in Fig. 2(a). Holes occupy these bands up to the Fermi energy, which is illustrated according to Dobrowolska \textit{et. al}\cite{Dobrowolska_Nature2012}.
%The valence band electrons and the Mn ions are antiferromagnetically coupled, so in the ground state their spins favor opposite alignment.\cite{Szczytko_PRB01,Wang_PRB08}
Conversely, the conduction bands are barely affected by the Mn spins due to their $s$-like symmetry and small $s$--$d$ coupling. 
%duction band can be ignored\cite{Kimel_PRL04}. 
According to the optical selection rule $\Delta J=\pm 1$, each circularly polarized branch of the incident probe light couples a unique electronic transition from one of the spin-polarized valence bands to the degenerate conduction band with opposing spins. 
Consequently, the origin of the static (no pump) MOKE signal below T$_{\text{C}}$ is caused by a difference in the refractive index experienced by each circular polarized branch of light, which is proportional to $\Delta_{MF}$.
%The size of $\Delta_{MF}$ depends on Temp and time after pump pulse....

In this paper we focus on the recovery of the non-equilibrium state following femtosecond pump excitation by using the fs-resolved MOKE signals to observe key information about the coupling between different reservoirs and their dissipations. 
Upon ultrafast pump excitation at 1.55eV, hot holes are photoexcited out of equilibrium giving rise to a blurred Fermi surface at such fast timescales.
%from the transient increase in the hole temperature at such fast timescales. 
Due to the strong exchange interaction between the holes and the Mn ions spins, the photoexcitation results in a transfer of angular momentum from ferromagnetically-ordered Mn to the hot holes, which quasi-instantaneously demagnetizes the Mn ions and creates an initial non-equilibrium spin-polarized hole state even with linearly-polarized light during femtosecond laser excitation at temperatures below T$_\text{C}$. The ultrafast demagnetization (amplitude change) in (III,Mn)V has been envisioned as the reverse of the Overhauser effect: the excited holes become dynamically spin polarized at the expense of the localized Mn spins \cite{Wang_PRL05, Cywinski_PRB07}. To study this non-equilibrium, photo-excited hole state, we set the probe to be degenerate with pump at 1.55eV, which directly couples to the transient carrier population. 

\begin{figure} 
\includegraphics[scale=.3]{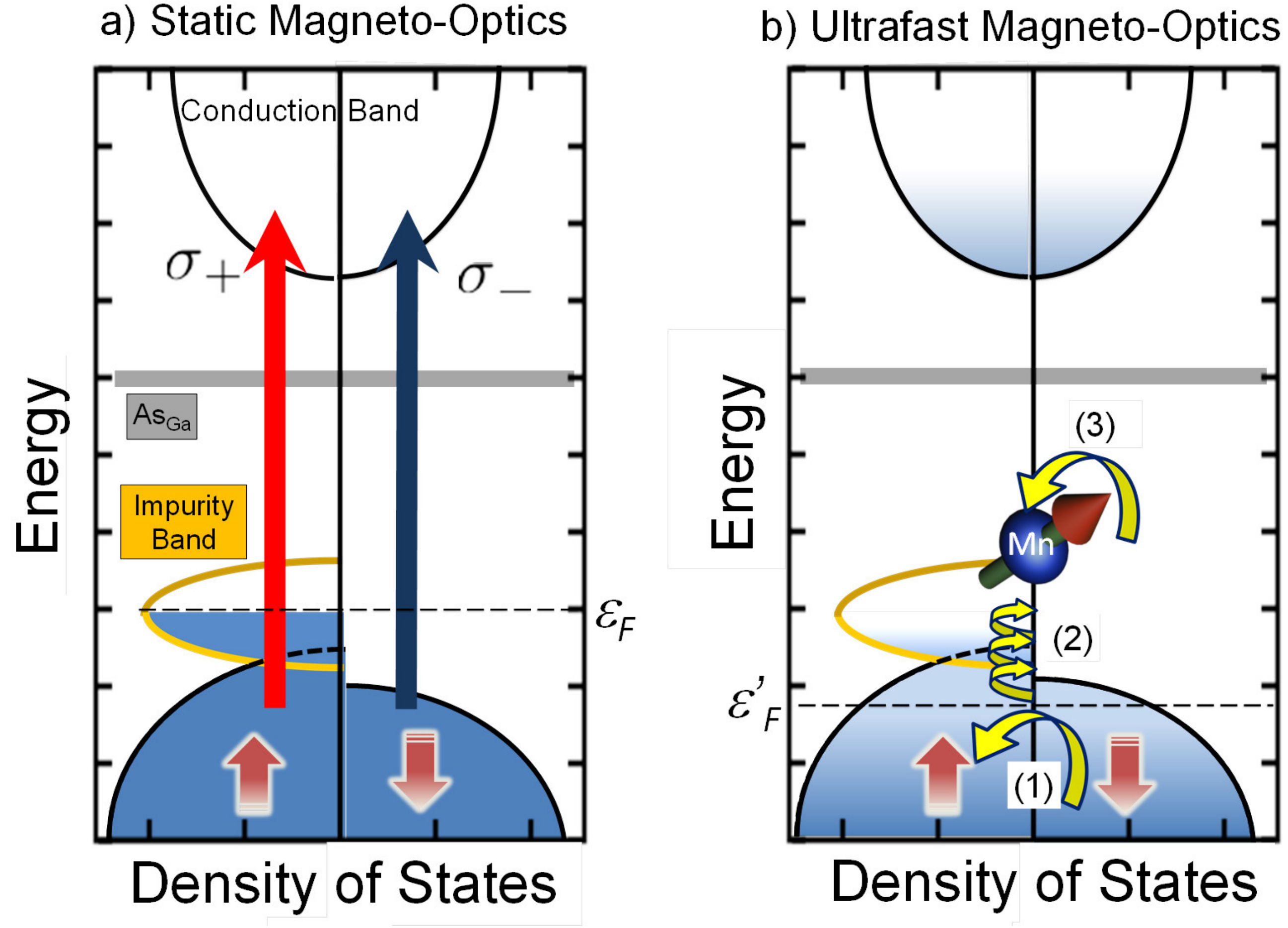} 
\caption{\label{fig2} Experimental schematics. (a) The static MOKE signal at $\hbar\omega=1.55eV$ arises from a difference in refractive index experienced by the two circular polarization modes of incident light. The valence band is split by $\Delta_{MF}$ which comes from the effective mean field of the ferromagnetic Mn ions. The $d$-like electronic orbitals of Mn form a dispersion-less, mid-gap impurity band (shown in gold) that hybridizes with the upper valence band \cite{Dobrowolska_Nature2012}. (b) Schematic of the three observed relaxation processes in this paper following photoinduced ultrafast demagnetization after pump pulse excitation at $\hbar\omega=1.55eV$: (1) hole spin relaxation via scattering between the spin split valence bands; (2) hot hole energy relaxation via phonon scattering within the valence and impurity band; (3) coherent Mn spin pressession and Gilbert damping.}
\end{figure}

Fig. 2(b) illustrates the three interconnected relaxation processes that contribute to the \emph{ultrafast} MOKE response following linearily polarized pump pulse excitation. The first process is the relaxation of transient spin-polarized holes illustrated as scattering between the spin split hole bands.  
%The non-equilibrium hole spin population is induced by the dynamical polarization of holes through $p$-$d$ exchange scattering with ferromagnetically-ordered Mn spins at temperatures below T$_{\text{C}}$. 
%This presents a new scheme to measure the hole spin lifetimes while more complicated methods have been used in semiconductors such as GaAs and Si with no static magnetic order to split the bands in the ground state, e.g., mid-infrared, circularly polarized pulses \cite{Hilton_PRL02}. 
Note that a spin-relaxation for the photoexcited conduction electrons does not occur in this case because the linearly polarized pump does not deposit any net angular momentum and, thereby, does not create any net electron spin, and additionally, there is no preferred spin orientation for the photoexcited conduction electrons to relax to.
%This spin relaxation allows the charge carriers to shed an amount of energy equal to the size of 
%the spin splitting $\Delta(T,t)$. It is written $\Delta(T,t)$ to emphasize that the spin splitting is 
%a function of both temperature and time.
%The second process is a blocking of transitions by high energy photoexcited holes added to the valence band, which diminishes the absorption coefficient for both circular polarizations and indicates the cooling of hot hole population towards zone center.
The second process is a relaxation of energy of the transient hot holes seen as cooling via phonon scattering towards zone center of the valence and impurity bands. Although there is no change of the spin polarization in this, this process contributes to the transient MOKE signals by altering the absorption coefficient for both circular polarizations, commonly referred to as dichroic bleaching in prior studies in ferromagnetic metals \cite{Zahn_APL2010, Koopmans_PRL00}. 
%As described in the \emph{Results and Analysis} section, we propose that the low-energy spin band preferentially decays via lower-energy/higher-momentum acoustic phonon scattering while the high-energy spin band decays more through relatively high-energy/low-momentum optical phonons. 
The third relaxation process is due to coherent Mn spin pressession and Gilbert damping via spin-lattice scattering.
Previous studies have only observed these last two processes and a measurement of the hole spin dynamics remains an open issue.
%Here, linearly polarized light was used as a pump to eliminate any influence from spin polarized conduction electrons.
%ost likely due to differing experimental conditions such as lower pump 
%fluence\cite{Kojima_PRB03,Kimel_PRL04,Wang_PRL07} and sample differences. 
%The above-mentioned relaxation processes occur on different timescales because of the different microscopic interactions invovled. Therefore, even though these processes are indistinguishable in static measurements, they can be separated in time when viewed at sufficiently short time scales (less than the period of relevent collective excitations).

\section{Results and Analysis}
A typical data trace of the ultrafast photoinduced MOKE rotation -$\Delta \theta_{k}$ at 4K is shown in Fig. 3(a). Note that the x-axis was split to show more clearly the short and long time dynamics. We observe an initial $\sim$40$\emph{fs}$ fast reduction (inset) 
%comparable to the 35fs pump pulse duration 
due to femtosecond demagnetization that gives the negative change in the ultrafast MOKE signals. Subsequently, a distinct three-step recovery of $\Delta \theta_{k}$ is marked with (1) a fast $\left(\tau_{S} \sim 200fs \right)$ decay, (2) a slow $\left(\tau_{E} \sim 2ps \right)$ decay, and (3) a periodic oscillation with a period of $\sim250$\emph{ps} superimposed on a much slower Gilbert relaxation.
%occurs through $p\!-\!d$ spin scattering between the photoexcited holes and localized Mn %ions.\cite{Wang_PRB08} This ultrafast demagnetization reduces the size of gap %$\Delta\left(T,t\right)$ in the valence band. Then, the hot, photoexcited holes in the valance %band and the electrons in the conduction band quickly thermalize during or immediately following %the pump pulse which results in a non-equilibrium distribution. 
%because the high energy spin band is now more thermally accesible. 
%results in further bleaching of the high-energy-spin valence band as they relax into the lower-enery-spin valence band and occurs within the optical pump pulse.\textbf{(cite somestuff here!!!!!!)}
%In the next sections we explain why our results suggest the long timescale dynamics originate from Mn spin and the short timescale dynamics come from photoexcited hole relaxation, in agreement with previous reports.\cite{Mitsumori_PRB04,Wang_PRL07} 

\begin{figure} 
\includegraphics[scale=0.57]{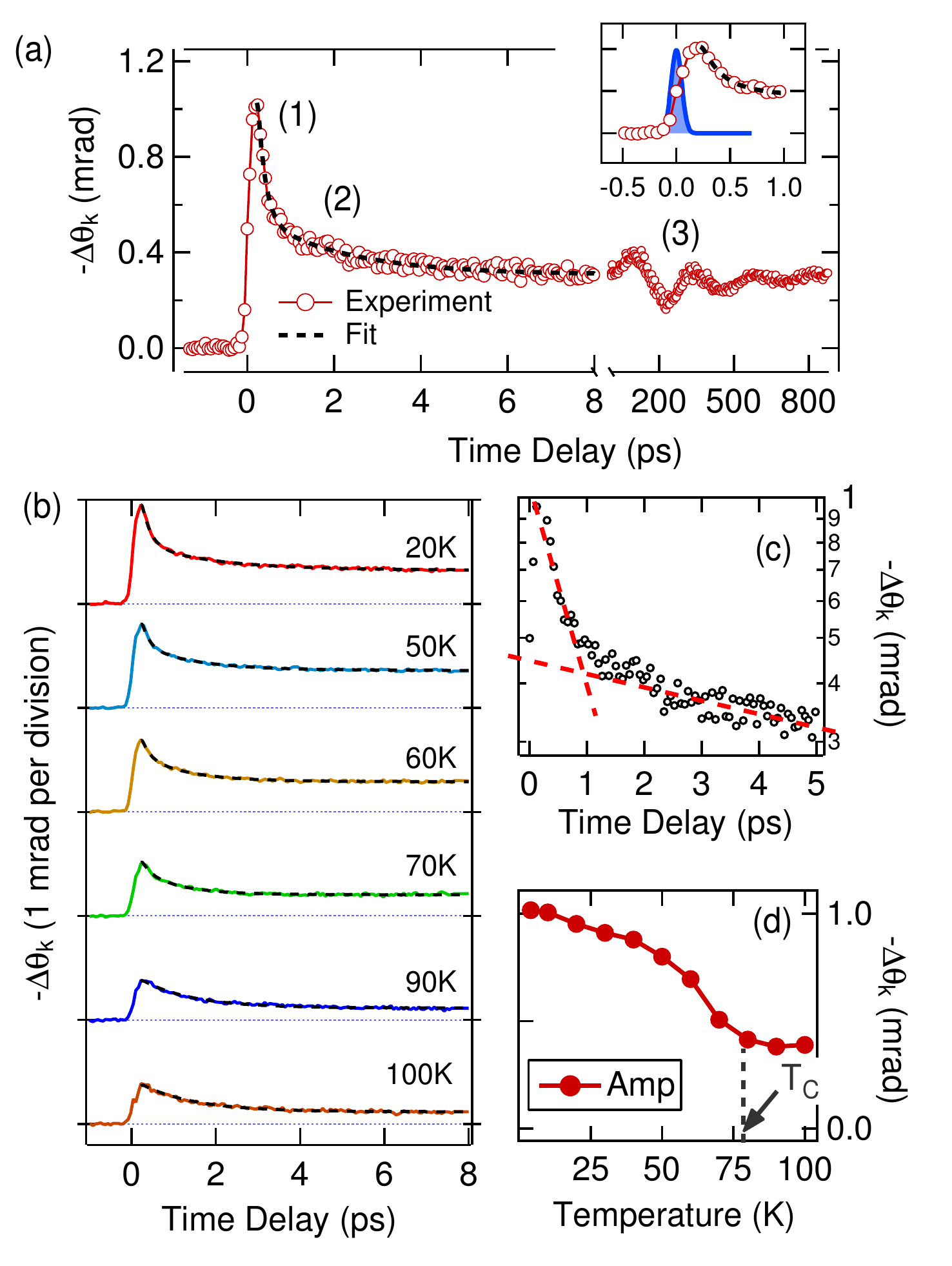} 
\caption{\label{fig3} (a) A representative trace of the low temperature data below T$_C$ with a split x-axis. A three-step recovery of the MOKE signal is observed following ultrafast negative rise (demagnetization): (1) fast $\sim$200\emph{fs} and (2) slow $\sim$2\emph{ps} time constants followed at longer timescales by (3) a periodic oscillation superimposed upon a much slower decay at 100s of ps timescale. Inset shows the cross-correlation of pump and probe (blue fill). (b) Temperature dependence of the MOKE dynamics during the first 8ps. Traces are vertically offset for clarity. Black dashed lines overlapping the data show the bi-exponential fitting, which is illustrated in a logarithmic-plot of the 4K trace in (c). Red dashed-lines in (c) mark each decay component. (d) Peak amplitude of the ultrafast photoinduced MOKE rotation. A magnetic field of B$_{\text{ext}} \sim 250\text{mT}$ was applied perpendicular to the sample and the pump fluence was $\sim$690$\mu$J$/$cm$^{2}$.}
\end{figure}
%Later, we present a brief analysis of the long time, Mn spin dynamics which show a................

\subsection{Short-time dynamics: hole spin and energy relaxation}
First, we turn the reader's attention to the partial recovery of the MOKE signal within the first few ps after photoexcitation, shown in Fig. 3(b). The fast dynamics largely reflects the relaxation of photoexcited holes, considering heavy $p$-doping in the ground state and previously discussed ultrafast relaxation processes.  
%, the fact that electron spin relaxation occurs on longer timescales of $\sim$30ps, and in addition to the fact that the near-band-gap photoexcitation excites electrons into the low lying states in the conduction band at the $\Gamma$-point where the energy of the photoexcited electrons is too small for significant inter-valley scattering \cite{Kimel_PRL04}. 
The bi-exponential relaxation dynamics at fast timescales are clearly seen in the logarithmic plot shown in Fig. 3(c), where two red, dashed lines overlay the T=4K trace (black, hollow dots) marking the relaxations processes. 

Ultrafast MOKE measurements for various temperatures from 20K up to 100K in Fig. 3(b) reveal salient features. %We initially cool down to far below the Curie temperature and then increase the temperature. 
Below T$_{\text{C}}$=77K, the data shows bi-exponential decay with $\tau_{S}$ and $\tau_{E}$ components, while only the slower $\tau_{E}$ component persists above T$_{\text{C}}$, which manifests as a single exponential decay. After crossing into the paramagnetic phase, there is no longer a macroscopic spin order and no dynamically polarized hole spins. In addition, since linearily polarized pumping does not deposit any angular momentum, one expects to only observe the hole spin relaxation below T$_{\text{C}}$. Therefore, we can attribute the slower component, $\tau_{E}$, to the hole energy relaxation. On the other hand, since the femtosecond component, $\tau_{S}$, disappears above T$_{C}$, we ascribe it to the hole spin relaxation (further support coming up).

%Based on the observation of two very different relaxation times, we thereby separate two of the photoexcited hole spin and energy relaxation processes in the time domain. 
%The hole-spin relaxation (200fs fast component) has not been observed before and is the primary focus of this study. 
We present, in the inset of Fig 3(a), the 4K trace alongside the pump-probe cross-correlation (blue fill) to illustrate that the fast $\tau_S$ relaxation is indeed temporally resolvable with our pulse duration. It is not due to a coherent interaction between the pump and probe beams, which is an interesting effect in itself however \cite{Hall_APL08,Han_APL2007}.

\begin{figure} 
\includegraphics[scale=.45]{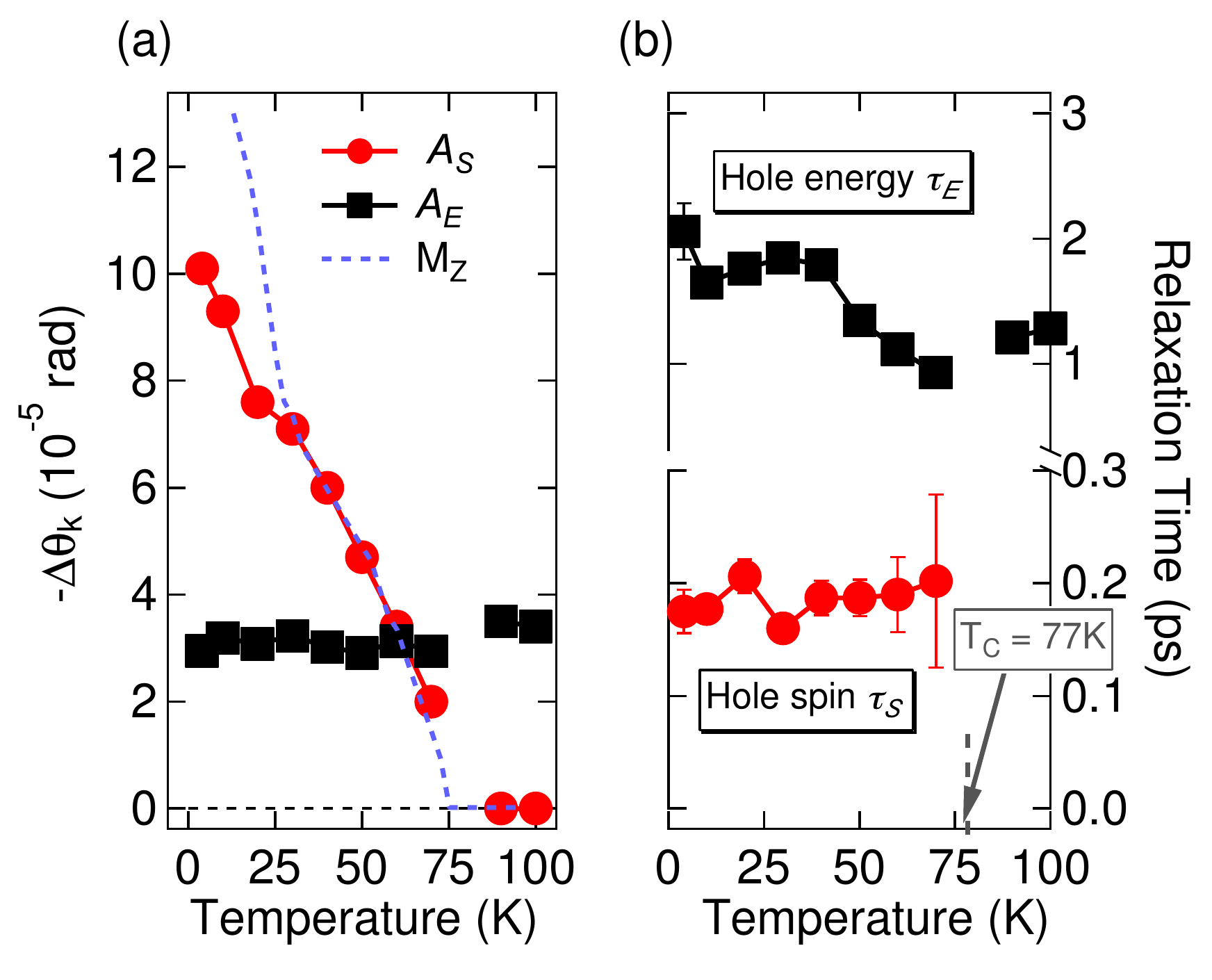}
\caption{\label{fig4} (a) The extracted transient amplitude for the hole-spin relaxation, $A_S$ (red), scales nicely with the static magnetization M$_Z$ (blue dashed line) for temperatures above $T_{R}$ while the amplitude of the hole-energy relaxation, $A_E$ (black), remains mostly constant with temperature. (b) Relaxation time constants as a function of temperature for $\tau_S$ (red) and $\tau_E$ components (black).}
\end{figure}

Furthermore, as shown in Fig. 3(d), the peak amplitude of the photoinduced $\Delta\theta_{k}$ signals decreases with increasing temperature up to the ferro- to paramagnetic transition, but the signal persists even above above T$_C$. These properties indicate that the ultrafast MOKE signal is measuring signals from both the ferromagnetic order and from the contributions of the non-equilibrium hole spin and charge populations as discussed in the prior section. In addition, the transient amplitude is constant above T$_C$, which corroborates our assignment of the slower component, $\tau_{E}$, to the hole energy relaxation.

%contributions to the ultrafast MOKE signal. The nonmagnetic contribution to the signal above Tc is mostly constant with temperature and further confirms that the identity of the $\tau_E$ component, is the non-magnetic, hole energy relaxation.

We fit the short time dynamics with a bi-exponential decay function convoluted with the probe pulse. The temperature dependent fittings are shown overlaying the data in Fig. 3(a,b) (black, dashed-lines). The bi-exponential fitting function is given by:
\begin{equation}
f\left(t\right)=A_{S}\ \text{exp}\left(-t/\tau_{S}\right)+A_{E}\ \text{exp}\left(-t/\tau_{E}\right)+A_{C}. \nonumber
\end{equation}
$\tau_{S}$ and $\tau_{E}$ are the characteristic decay time constants for the hole-spin and hole-energy relaxation processes, and $A_{S}$ and $A_{E}$ are their respective transient amplitudes. The third term represents the long-time Mn spin-relaxation and other non-magnetic contributions that are practically time-independent over the 8ps considered for the fitting, so these terms are represented as just a constant, $A_{C}$.
%A term representing the finite rise-time was originally implemented, but it was found to be tangled with the fast relaxation component due to similar timescales ($\tau_{rise} \sim$100fs). 
%Therefore, the finite rise-term was not used (and assumed constant) since the focus of this paper is on the relaxation processes following quasi-equilibrium, not on the rise time.

Fig. 4(a,b) shows the temperature dependence of the parameters acquired from the above fitting. The amplitude of the spin component A$_{S}$ (red dots) diminishes with increasing temperature and is absent above T$_C$, closely matching the static magnetization curve M$_Z$ (dashed blue line) shown for comparison in Fig. 4(a) \cite{M_Zscale}. The slight discrepancy at temperatures below T$_R$=30K likely originates from the spin reorientation and will be discussed in the next section. On the other hand, A$_{E}$ (black squares) stays constant in the whole temperature range and persists above T$_C$. These behaviors corroborate our conclusions of the magneto-optical nonlinearities in the MOKE signals; the magnetic-dependent amplitude A$_{S}$ measures the contribution of the spin-polarized hole population, which is generated via dynamical polarization transfer from the macroscopic magnetization of the Mn ions, and the corresponding decay time $\tau_S$ measures the hole spin relaxation times, while the non-magnetic amplitude A$_{E}$ measures the hot hole population independent of $T_C$ and the decay time $\tau_{E}$ comes from hot hole energy relaxation in the valence and Mn-impurity bands via phonon emission.

The fast hole-spin decay time $\tau_{S}$ (red solid dots) is mostly constant $\sim$160-200fs from 4K up to T$_C$ as shown in Fig. 4(b). These directly measured values are consistent with an upper bound value $\sim$200 fs estimated from the fs demagnetization experiment in (III,Mn)V ferromagnetic semiconductors \cite{Wang_PRL05}. These values are slightly larger than the $\sim$50--80fs predicted theoretically for GaMnAs \cite{Shen_PRB2012}. In these materials, the fast hole spin-relaxation can be attributed to spin-orbit coupling in the valence band mediated by ultrafast momentum scattering in the excited states that gives rise to a strongly fluctuating spin-orbit field acting on hole spins.  This mechanism results in very fast and temperature independent fs decay in the ferromagnetic phase, which is likely more efficient than any other relaxation mechanism that leads to temperature dependence as predicted \cite{Shen_PRB2012}. 

In addition, the hole energy relaxation time $\tau_{E}$ (black squares in Fig. 4b) stays constant until around $T_{R} \sim$ 30-40K, where it begins decreasing and exhibits a downward cusp near T$_C$, showing some dependence on the magnetization. This interesting behavior is still an open question, but it may originate from the strong band mixing between the $p$-type valance and $d$-Mn bands as well as the complicated spin-induced renormalization of the hole states. The revealed $\tau_{E}\sim$1-2ps relaxation time is consistent with the previously observed hole energy relaxation in GaMnAs and our results here further reveal its non-trivial temperature dependence that identifies the influence from ferromagnetism. Moreover, it has been previously established that the trapping of photoexcited electrons by crystal defects is also $\sim$ 1ps \cite{
%carrier1, carrier2, 
Mitsumori_PRB04, Zahn_APL2010, Kojima_PRB03}, similar to the $\tau_{E}$ relaxation time. However, the previous studies indicated that the photoexcited electrons were mostly decoupled from the magnetic system, but here, the temperature dependence of $\tau_E$ and its coupling to the magnetic system to some degree implies our $\tau_E$ component is not due to photoexcited electron relaxation. Our results point to the need for future studies and understanding of the hole-energy dynamics.

\subsection{Long-time dynamics: coherent Mn spin precession}
In this section, we briefly discuss the coherent oscillatory of Mn spins observed in the $\Delta \theta_k$ signals on the 100s of ps time scales, which has been reported in the literature \cite{Qi_PRB2009, Qi_APL2007, Rozkotova_APL2008, Scherbakov_PRL2010, Zhu_APL2009,  Hashimoto_PRL2008, Rozkotova_APL2008b}.  In order to isolate Mn dynamics without the contribution from hole dynamics, we use low pump fluence and a two-color method of ultrafast MOKE spectroscopy where we detune the pump to 3.1 eV to couple to energy levels far away from the fermi level at the $\Gamma$-point.
This allows us to avoid the overshoot behavior related to hole spin and energy dynamics as discussed in the prior section. 

At zero magnetic field and well below T$_C$, the initial magnetization direction is close to the sample easy axis [100] as shown in Fig. 1(a).  The magnetization dynamics after fs laser pulse excitations is presented in Fig. 5(a).  Here the main feature is the strong collective precession of the magnetization with a frequency of 4.2 GHz (inset). 
 The precessional motion is due to the photoexcited magnon as discussed in the literature. Our observed magnon frequency in the low field limit is well described analytically by the formula,
\begin{equation}
 \omega_{m} = \gamma \cdot \sqrt{H_{4\parallel}\cdot (H_{2\perp} + H_{4\parallel})} \nonumber
 \end{equation}
 where $H_{2\perp}$ and $H_{4\parallel}$ represent the effective uniaxial and cubic anisotropy fields, respectively. $H_{2\perp}$ and $H_{4\parallel}$ are defined in terms of anisotropy energies $K_{i}$ as $4 \pi M -\frac{2K_{2\perp}}{M}$ and $\frac{2K_{4\parallel}}{M}$. From the experimentally measured hard axis magnetization curve, shown in Fig. 1(b), we estimate $H_{2\perp}$ and $H_{4\parallel}$ to be 0.3T and 0.06T at 5K in our sample, which yields a magnon frequency $\omega_{m} = 4.12GHz$ in good agreement with the measured value. 
In addition, our results in Fig. 5(b) show that the GHz spin precession diminishes  
with increasing perpendicular $B$ field to 1T. 
As the magnetization aligns with the $B$--field  
along [001], $\Delta \theta_{k}$ reflects the 
fs changes in  magnetization  amplitude \cite{Wang_PRL05, Wang_JPC06} rather than changes in magnetization orientation $B$$\approx$0.

\begin{figure} 
\includegraphics[scale=0.11]{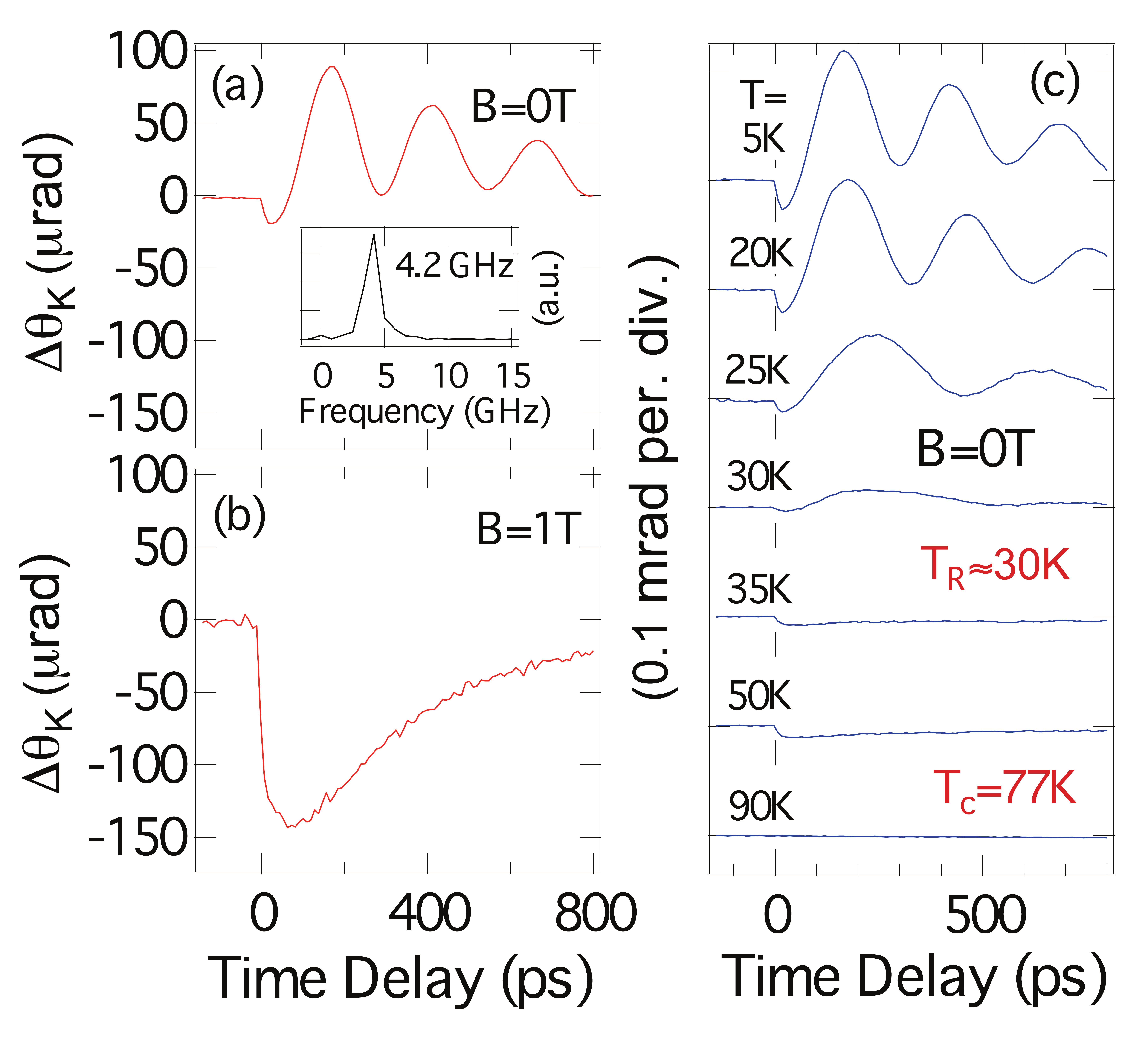} 
\caption{\label{fig6} (Color online) Ultrafast non-degenerate MOKE measurements at 0T (a) and 1T (b). T=4K and pump/probe photon energy is set to 1.55eV/3.1 eV. (c) The temperature dependence of ultrafast non-degenerate MOKE signals at various temperatures and traces are vertically offset for clarity. The pump fluence is $\sim$7$\mu$J$/$cm$^{2}$.}
\end{figure}

The GHz oscillation observed is triggered by the competition between uniaxial [1-10] and cubic [100] anisotropy fields. As shown in the Fig. 1(a), the magnetization in the sample aligns naturally along the dominant anisotropy field component, which at low temperatures is close to [100]. At these low temperatures, the thermal effects due to the pump pulse manifest as a strengthening of the uniaxial field [1-10] relative to the cubic field [100], equivalent to a change of the easy axis direction towards [1-10]. This induces precession of the magnetization along the new easy axis direction, consistent with the oscillations observed in the z direction, as shown in the 5K, 20K and 25K traces in Fig. 5(c). Around the easy-axis reorientation temperature T$_R \sim$30K, the uniaxial anisotropy field [1-10] becomes dominant, and the magnetization aligns along it. Thermal effects, strengthening this field, cannot induce precession of the magnetization anymore, as evident by the disappearance of the oscillations around T$_R$. In addition, the large anisotropy field associated with the [100] direction makes the magnetization vector tilt from the Z-axis below T$_R$, which explains the slight deviation between the static magnetization and A$_{S}$ in Fig. 4(a). It is interesting to note also that the ultrafast MOKE signals taken by the two-color scheme quickly diminish when approaching T$_C$, as seen in the high temperature traces shown in Fig. 5(c). This corroborates the assignment of the overshoot in degenerate MOKE signals to the hole dynamics instead of the Mn spin dynamics.

\section{Conclusions}
In summary, we have reported the previously-inaccessible non-equilibrium hole spin dynamics in ferromagnetic GaMnAs, measured with ultrafast MOKE spectroscopy. Below the Curie temperature T$_{\text{C}}$, an ultrafast linearly-polarized pump photoexcitation creates a non-equilibrium hole spin population via the dynamical polarization of holes through exchange scattering with ferromagnetically-ordered Mn spins. We then reveal the emergence of a new fs relaxation component below Tc, which we attribute to hole spin lifetime $\tau_{S}\sim$160-200 fs. This is distinguished in the time domain from the hole energy relaxation $\tau_{E}\sim$1-2 ps, and Mn spin precession on a 100 ps time scale. 
Our technique represents a new spectroscopy tool for studying the non-equilibrium hole spins in magnetically ordered materials, which could be important in understanding the various collective behaviors and macroscopic orders emerging in them. Additionally, our results have important implications for future applications in high speed spintronics using hole-mediated ferromagnetism. 

\begin{acknowledgments}
Research supported by the National Science Foundation (NSF) under Award $\#$DMR-1055352 (laser spectroscopy). Material
synthesis was supported by the National Science foundation Grant
DMR1400432. 
This work was also supported by the European Union's Seventh Framework Programme (FP7-
REGPOT-2012-2013-1) under grant agreement No. 316165 and by the EU Social Fund and National resources through the
THALES program NANOPHOS.

\end{acknowledgments}
% Create the reference section using BibTeX:
%\bibliography{bibfile}

%\bibliography{./refs_GaMnAs}

\begin{thebibliography}{99}
\bibitem[]{M_Zscale}
  \bibinfo{journal}{In Fig. 4(a), M$_Z$ along the (001) axis is scaled so that at T=4K, M$_Z$ relates to a static polar Kerr angle of $\Theta \sim$1.3mrad which is the requirement seen in Fig. 1(b) for an external magnetic field of $B=250mT$. }
\end{thebibliography}

\end{document}